\documentclass[aps,prl,twocolumn,superscriptaddress,floatfix,longbibliography]{revtex4-2}
\usepackage{setspace}
\usepackage{graphicx}
\graphicspath{{images/}}
\usepackage{caption}
\usepackage{subcaption}
\usepackage{amsmath,amssymb} 
\usepackage{xcolor} 
\usepackage{pgfplots}
\pgfplotsset{compat=1.7}
\usepackage{tikz}
\usetikzlibrary{arrows.meta, decorations.markings, positioning}
\usepackage{physics}
\usepackage{tensor}
\usepackage{threeparttable}
\usepackage{nicematrix}
\newcommand{\del}[2]{\frac{\partial #1}{\partial #2}}
\newcommand{\qav}[1]{\langle #1 \rangle}
\newcommand{\qp}[1]{\hat{#1}}
\newcommand{\vc}[1]{\textbf{#1}}
\newcommand{\bv}[1]{\boldsymbol{#1}}
\usepackage{listings}
\definecolor{codegreen}{rgb}{0,0.6,0}
\definecolor{codegray}{rgb}{0.5,0.5,0.5}
\definecolor{codepurple}{rgb}{0.58,0,0.82}
\definecolor{backcolour}{rgb}{0.97, 0.97, 0.97}
\lstdefinestyle{mystyle}{
	backgroundcolor=\color{backcolour},   
	commentstyle=\color{codegreen},
	keywordstyle=\color{magenta},
	numberstyle=\tiny\color{codegray},
	stringstyle=\color{codepurple},
	basicstyle=\ttfamily\footnotesize,
	breakatwhitespace=false,         
	breaklines=true,                 
	captionpos=t,                    
	keepspaces=true,                 
	numbers=left,                    
	numbersep=5pt,                  
	showspaces=false,                
	showstringspaces=false,
	showtabs=false,                  
	tabsize=2
}
\newif\ifptitle
\newif\ifpnumber
\newcounter{para}

\ptitletrue  
\pnumbertrue  

\newcommand{\mytitle}{Quantum statistics of single-mode radiation emitted by superradiant Dicke states}
\begin{document}


\title{\mytitle}

\author{Anirudh Yadav}
\affiliation{Department of Physics, 1150 University Avenue, University of Wisconsin-Madison, Madison, Wisconsin, 53706, USA}
\author{D. D. Yavuz}
\affiliation{Department of Physics, 1150 University Avenue, University of Wisconsin-Madison, Madison, Wisconsin, 53706, USA}
\date{\today}

\begin{abstract}
We study the quantum statistics of single-mode radiation emitted by an atomic ensemble when the ensemble is initially prepared in a superradiant Dicke state. We show that while the radiation is well approximated by the Glauber coherent state at early times in the evolution, the emission can be truly quantum at later times. In particular, one can observe a large amount of photon-number squeezing in the emission under certain conditions; even a Fock state can be produced. We discuss the quantum statistics of the emission for various parameters, including different initial conditions for the atomic ensemble. To obtain these results, we have developed a formalism where we are able to calculate the quantum statistics of the emission over long time-scales even when the number of atoms in the ensemble is quite large.

\end{abstract}
\maketitle
\textit{Introduction}---In his pioneering work more than 70 years ago, Dicke discussed superradiance in collective spontaneous emission from an ensemble of radiators \cite{Dicke}. Since this seminal work, superradiance has been experimentally observed in many different physical systems \cite{haroche}, including room-temperature atomic gasses \cite{feld,manassah}, ultracold ensembles \cite{bloch,an,gauthier,kuga,ions}, molecular systems \cite{molecules}, vacancy centers in solids \cite{diamond1,diamond2}, and superconducting circuits \cite{superconducting}. Over the last two decades, there has been renewed interest in collective spontaneous emission, in particular due to applications in quantum information science and quantum sensing \cite{ballantine,kimble,yelin,cirac,zanthier,agarwal,reitz,H_limit}.

While superradiance and collective spontaneous emission have a long history, it is well-known that this problem is theoretically quite difficult to study \cite{haroche}, primarily due to the exponentially-large dimension of the Hilbert space. Analytical solutions can only be found either for a very few number of atoms \cite{ph_stat_two}, or under certain assumptions that restrict the problem to a subspace of the Hilbert space \cite{francis,ph_stat_li}. Of particular importance, one of the outstanding open problems has been calculating the quantum statistics of the emitted photons under the conditions of superradiance when the number of atoms in the ensemble is large. The main results in this problem were derived more than 50 years ago, when Birula \cite{Birula} showed that emission from the maximally superradiant state (i.e., half of the atoms are in their excited level while the other half is in the ground level) is in a Glauber coherent state \cite{Glauber} at early times in the evolution. Since this result, very little progress has been made, and most recent papers focus on calculating the photon statistics for a low  number of atoms ($N \sim 10$), by numerically solving the exact density matrix in the evolution \cite{ph_stat_two,TC_few_atoms}.  

In this letter, for the first time to our knowledge, we develop a formalism to calculate emission from superradiant Dicke states at all times in the evolution even when the number of atoms in the ensemble can be quite large. Specifically, we will discuss two results: (1) We extend Birula's results for emission from the maximally superradiant state and find that at later times in the evolution, the photon state deviates significantly from the coherent state. In particular, under certain conditions the emitted field can approach an ideal photon-number Fock state. (2) We calculate the quantum statistics of the emission, not just from the maximally superradiant state, but also from states with different initial number of excited atoms. Qualitatively, the results are similar to the emission from the maximally superradiant state: at early times in the evolution, the photonic state is well approximated by the coherent state, while at later times there are deviations from the coherent state. Quantitatively, there are important differences in the emission when the initial number of excited atoms is varied. 

An important application of our results is to the generation of quantum light with statistics that are significantly different from the statistics of the coherent state. It is now well-understood that squeezed light, including Fock states, can be used to increase the quantum-limited sensitivity of optical measurements \cite{mandel,walls}. Such light sources also have applications in photonic based quantum computation, due to their non-Gaussian character \cite{dowling,pryde}.

\textit{Atomic Ensemble Interacting with Single-Mode Light}---We start with the following Hamiltonian that describes a single mode quantized light field interacting with an ensemble of $N$ two level atoms in the Dicke limit under the rotating wave approximation: 
\begin{equation}
	\qp{H}_{T}=\overbrace{\omega\qp{J}^{z}+\omega(\qp{a}^{\dagger}\qp{a}+1/2)}^{\qp{H}_{0}} \;+\; \overbrace{\Delta \qp{J}^{z}+(g^{*}\;\qp{a}\;\qp{J}^{+}+g\;\qp{a}^{\dagger}\;\qp{J}^{-})}^{\qp{H}}
\end{equation}
Here the quantities $\qp{a}$ and $\qp{a}^{\dagger}$ are the photon creation and annihilation operators, $\qp{J}^{z}=\sum_{i=1}^{N}\qp{\sigma}^{z}_{i}/2$, $\qp{J}^{\pm}=\sum_{i=1}^{N}\qp{\sigma}^{\pm}_{i}$  are the collective atomic operators, $\qp{\sigma}^{+}_{i}$ and $\qp{\sigma}^{-}_{i}$ are the atomic raising and lowering operators for the $a$th atom, $\Delta=\omega_{A}-\omega$ is the frequency detuning between the atomic transition energy and the photon mode, and the quantity $g=-\sqrt{\frac{\omega}{2\hbar\epsilon_0V}}d_{ge}\bv{\varepsilon}\cdot\bv{\varepsilon}_{A}$ is the coupling constant between atoms and the light field. This Hamiltonian is typically referred to as the Tavis-Cummings model \cite{TC_original} and it is known to have an exact solution for the eigenvalues, which can be evaluated explicitly for a low number of atoms \cite{TC_bgb,TC_few_atoms}. More recent studies on various aspects of this model are summarized in Refs.~\cite{TC_cavity_QED,TC_alg,TC_dis,TC_open,TC_beyond_mf,photon_blockade}. Since $\qp{H}_0$ commutes with $\qp{H}$, from here onward we only work with the interaction Hamiltonian $\qp{H}$. First, we consider diagonalizing the optical field in the coherent state basis (i.e., Glauber-Sudarshan $P$ representation) \cite{cstate_diag,mandel}: 
\begin{equation}
	\qp{H}=\int \frac{d^2z}{\pi}\ket{z}\qp{V}(z,z^{*})\bra{z} \quad ,
\end{equation}
where $d^{2}z=d[\text{Re}(z)]d[\text{Im}(z)]$, and $\qp{V}(z,z^{*})$ is \cite{cstate_diag}: 
\begin{align}
\nonumber &\qp{V}(z,z^{*})=e^{|z|^2}\int\frac{d^2\alpha}{\pi}\bra{-\alpha}\qp{H}\ket{\alpha}e^{|\alpha|^2}e^{\alpha^* z-\alpha z^*}\\
&=\Delta\qp{J}^{z}+
\left(g^{*}z\qp{J}^{+}+gz^{*}\qp{J}^{-}\right)\quad .
\end{align}
Next, to write the Schrödinger's equation, we need to choose a complete basis of states to represent the combined atom-photon state. The atomic operators $\qp{\sigma}^{z}_{i}/2$ and $\qp{\sigma}^{\pm}_{i}/2$ are spin-$1/2$ Pauli operators. The $N$-atom coupled operators $\qp{J}^{z}$, $\qp{J}^{x}$ and $\qp{J}^{y}$ hence satisfy the angular momentum algebra. Moreover, the Hamiltonian operator $\qp{V}$ satisfies the following property, $[\qp{V},\qp{J}^{2}_{1}]=[\qp{V},\qp{J}^{2}_{2}]=[\qp{V},\qp{J}^{2}_{12}]=[\qp{V},\qp{J}^{2}_{123}]\cdots=[\qp{V},\qp{J}^{2}]=0$, where, 
\begin{equation}
	\qp{\vc{J}}_{12\dots r}=\sum_{i=1}^{r}\frac{\qp{\bv{\sigma}}_{r}}{2}\;,\quad \qp{J}^{2}_{12\dots r}=\qp{\vc{J}}_{12\dots r}\cdot\qp{\vc{J}}_{12\dots r}\quad ,
\end{equation}
for, $2\le r\le N-1$. Therefore, we choose the coupled states $\ket{\pi j,\mu}$ as the basis for the atomic system. Here, $\pi$ is one configuration from all the possible coupling configurations $\{j_1,j_2,(j_{12}),\dots,(j_{12\dots N-1}),j_{N}\}$. The states $\ket{\pi j,\mu}$ are simultaneous eigenstates of the Hermitian operators $\qp{J}^{2}_{12\dots r}$, $\qp{J}^{2}$, and $\qp{J}^{z}$ with the following eigenvalues:
\begin{subequations}
\label{e}
\begin{align}
	\label{ea}
	\qp{J}^{2}_{12\dots r}\ket{\pi j,\mu}&=j_{12\dots r}(j_{12\dots r}+1)\ket{\pi j,\mu}\quad ,\\
	\label{eb}
	\qp{J}^{2}\ket{\pi j,\mu}&=j(j+1)\ket{\pi j,\mu}\quad ,\\
	\label{ec}
	\qp{J}^{z}\ket{\pi j,\mu}&=\mu\ket{\pi j,\mu}\quad ,
\end{align}
\end{subequations}
and thus form a complete orthonormal set of basis states for the atomic system. Hence, we represent the general state $\ket{\psi(t)}$ in the combined atom-photon basis $\ket{\pi j,\mu}\ket{z}$ as follows,
\begin{equation}
\ket{\psi(t)}=\sum_{\pi,j,\mu}\int\frac{d^{2}z}{\pi}\ket{\pi j,\mu}\ket{z}\overbrace{\bra{z}\braket{\pi j,\mu}{\psi(t)}}^{e^{-|z|^{2}/2}\psi^{\mu}_{\pi j}(t,z^{*})}\;.
\end{equation}
The Schrödinger's equation, 
\begin{equation}
    i\frac{d}{dt}\ket{\psi(t)}=\qp{H}\ket{\psi(t)} \quad ,
\end{equation}
in this basis takes the form (Einstein summation convention is employed):
\begin{equation}
	\label{se1}
    \sum_{\pi,j,\mu}\int\frac{d^{2}z}{\pi}\ket{\pi j,\mu}\ket{z}\left\{i\dot{\psi}^{\mu}_{\pi j}-\tensor{(V_{\pi j})}{^\mu_\sigma}\psi^{\sigma}_{\pi j}\right\}=0 \quad .
\end{equation}
Since the coherent states $\ket{z}$ are not independent, the expression inside the curly brackets in Eq.~(\ref{se1}) is not the evolution equation. To proceed further we project Eq.~(\ref{se1}) onto another coherent state $\bra{\xi}$ and integrate out $z$. This integration can be performed using reproducing properties of the coherent state basis, which is discussed in detail in the Appendix A of the End Matter section. The end result is the following evolution equation:
\begin{equation}
\label{se2}
    i\dot{\psi}^{\mu}_{\pi j}=(\Delta \tensor{(J^{z})}{^{\mu}_{\sigma}}+g^{*}\tensor{(J^{+})}{^{\mu}_{\sigma}}\partial_{\xi^{*}}+g\tensor{(J^{-})}{^{\mu}_{\sigma}}\xi^{*})\psi^{\sigma}_{\pi j}\; .
\end{equation}
This form of the evolution equation may deceptively suggest that the state does not evolve unitarily. However, we prove unitary evolution in the Appendix B of the End Matter section and provide an exact unitary solution to Eq.~(\ref{se2}). Solution to Eq.~(\ref{se2}) can formally be written as the action of the following propagator: 
\begin{equation}
\label{a1}
    \psi^{\mu}_{\pi j}(t,\xi^{*})=\tensor{[\hat{K}_{\pi j}(\partial_{\xi^{*}},\xi^{*})]}{^{\mu}_{\sigma}}\psi^{\sigma}_{\pi j}(t=0,\xi^{*}) \quad . 
\end{equation}
We require to calculate the matrix elements of the above propagator. In order to find that, we use the polynomial representation $\Phi_{\pi j,\mu}(u_{0},u_{1})$ for states $\ket{\pi j,\mu}$ for some fixed configuration $\pi$ and $j$, which is due to Wigner [\cite{Wigner}, p.\,163, Eq.~ (15.18)]:
\begin{equation}
    \Phi_{\pi j,\mu}(u_0,u_1)=\frac{u^{j-\mu}_{0}u^{j+\mu}_{1}}{\sqrt{(j-\mu)!(j+\mu)!}} \quad . 
\end{equation}
Here, the quantities $u_0$ and $u_1$ are symbolic variables denoting the ground and excited states respectively. The main novelty of this polynomial representation is that the collective angular momentum operators $\qp{J}^{z}$ and $\qp{J}^{\pm}$ can be expressed as differential operators:
\begin{align}
    \qp{J}^{z}=\frac{1}{2}\left(u_{1}\partial_{u_1}-u_{0}\partial_{u_0}\right),\;
    \qp{J}^{+}=u_{1}\partial_{u_0},\; \qp{J}^{-}=u_{0}\partial_{u_1}\; .
\end{align}
\begin{figure*}[t]
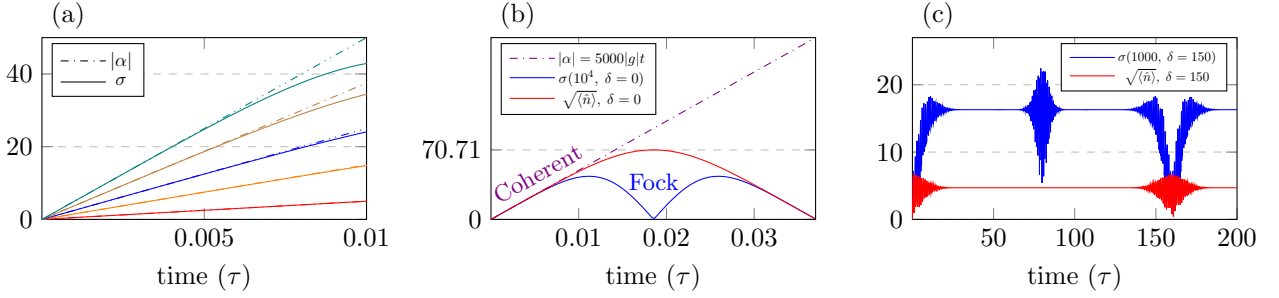

	\centering
	\begin{subfigure}{0.329\textwidth}

	\end{subfigure}
	\caption{In (a) we compare the standard deviation  of the photon number distribution, $\sigma(N)$ (solid lines), obtained from Eq.~(\ref{dm}) with the coherent state amplitude $|\alpha|=\tau N/2$ (broken lines) predicted by Birula in \cite{Birula} at early times $\tau\ll1/\sqrt{N/2}$. The number of atoms is $N=1000,3000,5000$ $7500$, and $10^{4}$ from bottom to top respectively. Plot (b) demonstrates strong photon-number squeezing ($Q_{\min}\approx -0.9998$) and shows that states close to the Fock state $\ket{N/2}$ at later times can be produced (i.e., the photon statistics deviates significantly from coherent Poisson distribution). In plot (c) we see that for large enough values of the detuning $\delta=\Delta/|g|$, the photon statistics exhibit different phases over large time scales.}
	\label{fig1}
\end{figure*}
The matrix elements of $\qp{K}$ can now be calculated using the following procedure. For any arbitrary function of the coherent state variable, $f(\xi^{*})$, we define the time-evolved state as $\qp{K}f(\xi^{*})\ket{\pi j,\mu}=\ket{G(t)}\equiv G(t,\xi^{*},u_{0},u_{1})$. Using the definition of the propagator, this time-evolved state is:
\begin{equation}
    G(t,\xi^{*},u_{0},u_{1})=\exp(-it\qp{V}(\partial_{\xi^{*}},\xi^{*}))f(\xi^{*})\Phi_{\pi j,\mu}(u_0,u_1)\; .
\end{equation}
Differentiating with respect to time once, we obtain the following evolution equation:
\begin{equation}
\label{se3}
    i\del{G}{t}=\frac{\Delta}{2}\left(u_{1}\del{G}{u_1}-u_{0}\del{G}{u_0}\right)+g^{*}u_1\frac{\partial^2G}{\partial u_0\partial \xi^{*}}+g\xi^{*}u_0\frac{\partial G}{\partial u_1}\; . 
\end{equation}
As a result of Wigner's polynomial representation of Eq.~(11), the multiple partial differential equations (PDEs) in Eq.~(\ref{se2}) reduce to a single PDE in time, $t$, and in canonical coordinates, $u_0$ and $u_1$, as shown in Eq.~(\ref{se3}). To solve this PDE, we observe that in Eq.~(\ref{se3}) $g$ and $\Delta$ are time independent and in the last two terms, every power of $u_{0}$ is accompanied with $\xi^{*}$ while every power of $u_{1}$ with $e^{-i\phi} (\phi=\arg(g)$), furthermore $[\qp{K},\qp{J}^{2}]=0$. Based on these observations, we propose the following ansatz:
\begin{equation}
    G=\sum_{n=-\infty}^{\infty}\sum_{q=-j}^{j}G^{n}_{q}(t)\frac{(\xi^{*})^{n}\Phi_{\pi j,q}(\xi^{*}u_{0},e^{-i\phi}u_{1})}{\sqrt{(n+j-q)!}} \quad . 
\end{equation}
Substituting the ansatz in Eq.~(\ref{se3}) and simplifying we obtain the following matrix equation,
\begin{equation}
	i\dot{G}^{n}_{q}=|g|\sum_{q'=-j}^{j}M^{j,n}_{qq'}\;G^{n}_{q'}
\end{equation}
\begin{figure}[b]
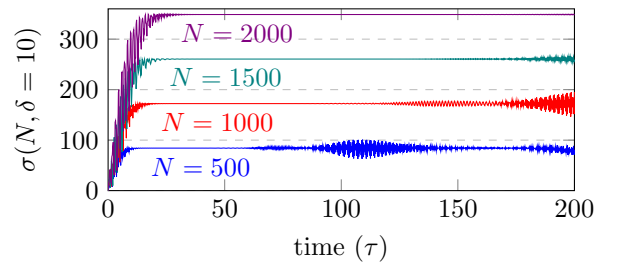


\caption{The standard deviation of the photon number distribution $\sigma$ for various number of atoms $N$ at $\delta=\Delta/|g|=10$. The time duration between different phases of the distribution increases with increasing $N$. We observe a similar trend for the average number of photons in the emitted field as a function of time.}
	\label{fig2}
\end{figure}
where, 
\begin{align}
    \nonumber &M^{j,n}_{qq'}=q'\frac{\Delta}{|g|}\;\delta_{qq'}\\
    \nonumber &+\sqrt{(j-q')(j+q'+1)(j-q'+n)}\;\;\delta_{q-1,q'}\\
    &+\sqrt{(j+q')(j-q'+1)(j-q'+1+n)}\;\;\delta_{q+1,q'}\quad .
\end{align}
is a symmetric matrix that acts as an the effective Hamiltonian. Imposing the initial condition for the combined photon-atom wavefunction,
\begin{equation}
	\ket{G(t=0)} = f(\xi^{*})\ket{\pi j,\mu}=\sum_{\nu=0}^{\infty}\frac{f^{(\nu)}(0)}{\nu!}(\xi^{*})^{\nu}\ket{\pi j,\mu}\; ,
\end{equation}
\begin{figure*}[t]
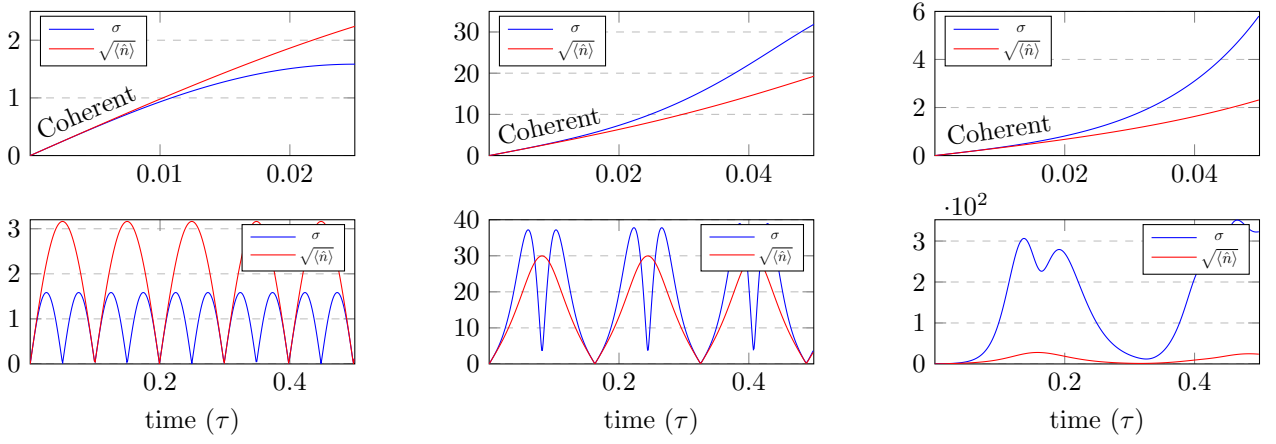

	\centering
	\begin{subfigure}{0.329\textwidth}

		\caption{$N=1000,\; \delta=0,\; M=-490$.}
	\end{subfigure}
	\begin{subfigure}{0.329\textwidth}
		%
		\caption{$N=1000,\; \delta=0,\; M=400$.}
	\end{subfigure}
	\begin{subfigure}{0.329\textwidth}
		%
		\caption{$N=1000,\; \delta=0,\; M=500$.}
	\end{subfigure}
	\caption{Photon statistics of emission from different Dicke states for  $N=1000$ atoms and assuming on-resonance condition, $\delta = 0$. In (a) only $10$ atoms initially start in the excited level and the emitted radiation is significantly squeezed. In (b) $900$ atoms are excited and the amount of squeezing is less compared to (a). When all atoms start in their excited level in (c), mean photon number rapidly expands and the emitted radiation is predominantly super-Poissonian.}
	\label{fig3}
\end{figure*}
we infer that, 
\begin{equation}
	G^{n}_{q'}(t=0)=\delta^{n,\mu+\nu-j}\delta_{q'\mu}\frac{f^{(\nu)}(0)}{\sqrt{\nu!}}e^{i(j+\mu)\phi} \quad ,
\end{equation}
for $\nu=\{0,1,2,\dots\}$. Finally, since $M^{j,n}_{qq'}$ is time independent we find, 
\begin{align}
	\label{sol2}
	\nonumber &\tensor{(\hat{K}_{\pi j})}{^q_\mu}f(\xi^{*})\\
	&=e^{i(\mu-q)\phi}\sum_{\nu=0}^{\infty}\frac{f^{(\nu)}(0)}{\sqrt{\nu!}}U^{j,\nu+\mu-j}_{q\mu}\frac{(\xi^{*})^{\nu+\mu-q}}{\sqrt{(\nu+\mu-q)!}} \quad . 
\end{align}
where,
\begin{equation}
	U^{j,n}_{qq'}=\exp(-i|g|t\;M^{j,n})_{qq'}\quad .
\end{equation}

\textit{Photon Statistics}---We now consider the photon statistics of the emitted radiation when the ensemble starts initially in a Dicke state and the photon field starts in vacuum: i.e, the initial condition for the system is $\ket{\psi(0)}=\ket{J,M}\ket{0}$. The evolution of this state can be computed using Eq.~(\ref{sol2}) with $\mu=M$ and $f(\xi^{*})=1$: 
\begin{equation} 
	\psi^{\mu}_{\pi j}(t=0,\xi^{*})=\delta_{\pi j,J}{\delta^{\mu}}_{M}\; 1 \quad , 
\end{equation}
which then gives 
\begin{align}
	\nonumber \psi^{q}_{\pi j}(t,\xi^{*})&=\delta_{\pi j,J}{(K_{J})^{q}}_{M}1\\
	\label{ev1}&=\delta_{\pi j,J}e^{i(M-q)\phi}U^{J,M-J}_{q,M}\frac{(\xi^{*})^{M-q}}{\sqrt{(M-q)!}}\; .
\end{align}
Since, $M-J<0$, the matrix $M^{J,M-J}$ has the following form,
\begin{equation}
	\begin{array}{cc}
	& 
	\begin{array}{c|c}
	-J\cdots\cdot\cdot\cdot\cdot M &\, M+1 \cdots J\\
	\hline 
	\end{array}\\[2pt]
	M^{J,M-J}=
	&
	\left(\begin{array}{cccccccc|ccccccccc}
		&&&&H'&&&&&&&0&&&\\
		\hline
        &&&&0&&&&&&&i\,\Gamma&&&
	\end{array}\right)
	\end{array} \quad . 
\end{equation}
where, $H'$ and $\Gamma$ are $J+M+1$ and $J-M$ dimensional real symmetric square matrices. The index $q$ in Eq.~(\ref{ev1}) thus belongs to the set $\{-J,-J+1,\dots,M\}$, and the total number of excitations (photonic and atomic) is conserved throughout the evolution, $[\qp{H}_{0},\qp{H}]=[\qp{H}_{0},\qp{K}]=0$. The final state using Eq.~(9) is then:
\begin{equation}
	\ket{\Psi(t)}=e^{-i\qp{H}_{0}t}\sum_{q=-J}^{M}e^{i(M-q)\phi}U^{J,M-J}_{q,M}\ket{J,q}\ket{M-q}\;.
\end{equation}
Tracing out the atomic degrees of freedom, we obtain the reduced density matrix for the photon mode:
\begin{equation}
	\label{dm}
	\qp{\rho}_{o}(\tau)=\sum_{n=0}^{J+M}|U^{J,M-J}_{M-n,M}(\tau)|^{2}\op{n},\quad \tau=|g|t 
\end{equation}
We first consider the case when the atomic ensemble initially starts at a maximally superradiant state, $M=0$ ($N$ is assumed to be even and $N\gg 1$). For this case we recover the results found by Birula \cite{Birula} for small times $\tau\ll 1/\sqrt{N/2}$, which is shown numerically in Fig.~\ref{fig1}(a). The plot shows that the emitted radiation is in a coherent state for small times with mean photon count scaling with $N$ as $|\alpha|^{2}=\tau^{2}N^{2}/4$, and the standard deviation equal to the square-root of the mean $\sigma = |\alpha|$. The coherent state is a good approximation during the times $\tau\ll 1/\sqrt{N/2}$, and at later times in the time evolution, we observe significant deviation from the coherent state, which is shown in Fig.~\ref{fig1}(b). In particular, at certain points in the time evolution, the state evolves almost completely into a Fock state with an average of $N/2$ photons. At even later times in the evolution, we observe oscillations between the different phases of the photon state distribution for large enough detuning $\delta=\Delta/|g|$ as shown in Fig.~\ref{fig1}(c). Furthermore, we observe that the duration between these different phases increases with the increase in the number of atoms in the ensemble. This is shown in Fig.~\ref{fig2}. \\

\begin{figure}[t]
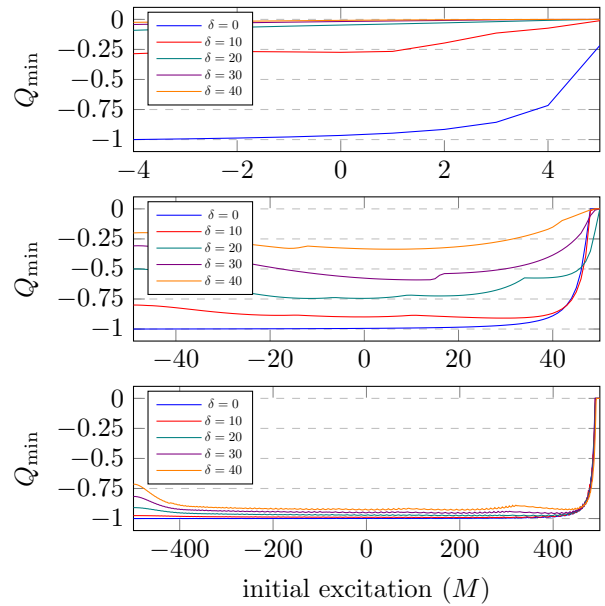

\begin{subfigure}{\linewidth}

\end{subfigure}
    \caption{Mean photon number expansion trend for $N=10,100$ and $1000$ atoms, respectively. The plots show the Mandel-Q parameter as a function of the number of initial excitation in the ensemble, $M$. We observe that for larger number of atoms the emitted radiation is more significantly squeezed even for large detunings when $M$ is small. The rapid expansion of the mean photon number and super-Poissonian behavior is mostly observed when at least $90\%$ of the atoms are initially excited.}
    \label{fig4}
\end{figure}

Next we study photon statistics when the atomic ensemble starts in more general Dicke states with $M\neq 0$. Computing the early time statistics of the photon field using the density matrix found in Eq.~(\ref{dm}) suggests that state starts out as a coherent state similar to the case for the maximally superradiant state: this is shown in the first row of Fig.~\ref{fig3}(a)-(c). The later time dynamics plotted in the second row of Fig.~\ref{fig3}(a)-(c) show that the state either squeezes in photon-number to almost a Fock state $\ket{N/2+M}$ or it experiences a rapid increase in the mean photon count reaching a maximum of $N/2+M$, depending on how large the initial number of excited atoms are. To quantify the boundary between these two qualitatively different behavior (photon-number squeezing versus rapid expansion in the mean photon number), we plot the minimum observed normalized variance/Mandel-Q parameter in the first cycle of evolution,
\begin{equation}
    Q_{\min}=\min_{0<\tau<1}\frac{\sigma^{2}(\tau)-\qav{\qp{n}}(\tau)}{\qav{\qp{n}}(\tau)} \quad , 
\end{equation}
as a function of $-J<M\le J$ for various values of the reduced detuning $\delta=\Delta/|g|$ and $N$ in Fig.~\ref{fig4}.\\

\textit{Conclusions and Discussion}---In conclusion, we have developed a formalism to calculate emission from superradiant Dicke states at all times for ensembles with a large number of atoms. Our formalism is valid not just for the maximally superradiant state, but also for Dicke states with a different number of excited atoms. We have confirmed that at early times in the evolution, the emitted field is well approximated by the Glauber coherent state. However, at later times, the photon state deviates significantly from the coherent state and under certain conditions, the emitted field can approach a Fock state. As we mentioned above, an important application of our results is to the generation of quantum light with statistics that are significantly different from the statistics of the coherent state \cite{mandel,walls,dowling,pryde}. 

We have found that the formalism above can be extended to calculate the emitted fields even when the ensemble is not in the Dicke limit (i.e., when the size of the ensemble is larger than the radiation wavelength) and/or when there are multiple modes of the field coupled to the same atomic ensemble. These results will be reported in a future publication. Experimental verification of some of the predictions of these results would be very interesting. While most experiments on photon statistics in collective coupling have been performed in free-space \cite{free_space_stat,thermal_atoms,spin_light_sq,spin_sq}, recent experiments have also started to explore emission into a single mode by placing  ultracold ensembles in a high-finesse cavity \cite{yan}. Extending these experiments to a large number of atoms, one can investigate the quantum statistics of emission into the cavity mode. The mode-field that leaks through the output coupler can be directly detected on a photon counter and the photon-number statistics can be studied. Alternatively, the leaked wave can be sent into a homodyne detection set-up using beating with a local oscillator and fluctuations in the quadratures can be measured. One immediate goal would be to detect the deviation from the coherent state and the generation of near-Fock states, as, for example shown in Fig.~1(b). 
\\

\textit{Acknowledgments}---We thank Prateek Gupta for helpful discussions. This work was supported by the National Science Foundation (NSF) Grant No. 2016136 for the QLCI center Hybrid Quantum Architectures and Networks (HQAN) and NSF Grant No. 2308818 from the AMO-Experiment program.

\section{End Matter}
\subsection{Appendix A: Reproducing property of coherent state distributions}
With the help of the integral below:
\begin{align}
\nonumber &\int\frac{d^2\alpha}{\pi}e^{v^*\alpha-|\alpha|^2}\alpha^p(\alpha^*)^q\\
\nonumber &=(\partial_{v^*})^p\int\frac{d^2\alpha}{\pi}e^{v^*\alpha-|\alpha|^2}(\alpha^*)^q\\
\label{int1}&=(\partial_{v^*})^p(v^*)^q \quad ,
\end{align}
we can evaluate the integral of a function $F(\alpha,\alpha^*)$ which is analytic independently in the variables $\alpha$ and $\alpha^*$ with the same kernel $e^{v^*\alpha-|\alpha|^2}$: 
\begin{align}
\nonumber &\int\frac{d^2\alpha}{\pi}e^{v^*\alpha-|\alpha|^2}F(\alpha,\alpha^{*})\\
\label{int2}&=\sum_{p=0}^{\infty}\sum_{q=0}^{\infty}\frac{1}{p!}\frac{1}{q!}\frac{\partial^{p+q}F}{\partial^{p}\alpha\;\partial^{q}\alpha^{*}}\Bigg\rvert_{(0,0)}(\partial_{v^{*}})^{p}(v^{*})^{q}\doteq\hat{F}(\partial_{v^{*}},v^{*})\; .
\end{align}
Therefore, this provides us with a differential representation for an integral operator, i.e.~given a state $g(\alpha^{*})$ we have the following due to Eq.~(\ref{int2}):
\begin{align}
\nonumber &\int \frac{d^2\alpha}{\pi}e^{v^*\alpha-|\alpha|^2}F(\alpha,\alpha^{*})g(\alpha^{*})\\
\nonumber
&=\sum_{p=0}^{\infty}\sum_{q=0}^{\infty}\frac{1}{p!}\frac{1}{q!}\frac{\partial^{p+q}F}{\partial^{p}\alpha\;\partial^{q}\alpha^{*}}\Bigg\rvert_{(0,0)}(\partial_{v^{*}})^{p}\left\{(v^{*})^{q}g(v^{*})\right\}\\
&\doteq \hat{F}(\partial_{v^{*}},v^{*})g(v^{*})\; .
\end{align}

\subsection{Appendix B: Proof of unitary evolution}
For a state which is normalized at all times $t$, we require:
\begin{equation}
    \braket{\psi(t)}=1\quad .
\end{equation}
In the basis $\ket{\pi j,\mu}\ket{z}$ the above can be re-expressed as follows:
\begin{align}
    &\sum_{\pi,j,\mu}\int\frac{d^{2}z}{\pi}\braket{\psi(t)}{\pi j,\mu}\ket{z}\bra{\pi;j,\mu}\braket{z}{\psi(t)}\\
    \label{cond1}&=\sum_{\pi,j}\int\frac{d^{2}z}{\pi}e^{-|z|^{2}}\psi^{*}_{\pi j\;\mu}(t,z)\psi^{\mu}_{\pi j}(t,z^{*})=1\quad .
\end{align}
States satisfying the evolution equation (\ref{se1}) can be shown to also satisfy the normalization condition in Eq.~(\ref{cond1}). The proof proceeds as follows (for brevity we suppress dependence $\pi, j$), the evolution equation and its conjugate are:
\begin{align}
    i\dot{\psi}^{\mu}&={V^{\mu}}_{\nu}\psi^{\nu}\\
    -i\dot{\psi}^{*}_{\mu}&={(V^{\dagger})^{\nu}}_{\mu}\psi^{*}_{\nu}
\end{align}
dotting the first with $\psi^{*}_{\mu}$ and the second with $\psi^{\mu}$, then subtracting, we obtain:
\begin{align}
    \nonumber &i\del{}{t}(\psi^{*}_{\mu}\psi^{\mu})={V^{\mu}}_{\nu}\psi^{\nu}\psi^{*}_{\mu}-{(V^{\dagger})^{\nu}}_{\mu}\psi^{*}_{\nu}\psi^{\mu}\\
    \nonumber &=[\Delta {(J^{z})^{\mu}}_{\nu}\psi^{\nu}\psi^{*}_{\mu}+g^{*}{(J^{+})^{\mu}}_{\nu}(\partial_{\xi^{*}}\psi^{\nu})\psi^{*}_{\mu}+g\xi^{*}{(J^{-})^{\mu}}_{\nu}\psi^{\nu}\psi^{*}_{\mu}]\\
    \nonumber &-[\Delta {(J^{z})^{\nu}}_{\mu}\psi^{*}_{\nu}\psi^{\mu}+g{(J^{-})^{\nu}}_{\mu}(\partial_{\xi}\psi^{*}_{\nu})\psi^{\mu}+g^{*}\xi{(J^{+})^{\nu}}_{\mu}\psi^{*}_{\nu}\psi^{\mu}]\\
    \nonumber &=g^{*}{(J^{+})^{\mu}}_{\nu}\left[\left\{\partial_{\xi^{*}}\psi^{\nu}(t,\xi^{*})\right\}\psi^{*}_{\mu}(t,\xi)-\xi\psi^{\nu}(t,\xi^{*})\psi^{*}_{\mu}(t,\xi)\right]\\
    &+g{(J^{-})^{\mu}}_{\nu}\left[\xi^{*}\psi^{\nu}(t,\xi^{*})\psi^{*}_{\mu}(t,\xi)-\psi^{\nu}(t,\xi^{*})\left\{\partial_{\xi}\psi^{*}_{\mu}(t,\xi)\right\}\right]
\end{align}
Finally, multiplying with the Gaussian weight $e^{-|\xi|^{2}}$, integrating with measure $d^{2}\xi/\pi$ and summing over all configurations $\pi,j$, the equation transforms to,
\begin{align}
    \nonumber &i\frac{d}{dt}\braket{\psi(t)}\\
    \nonumber &=\sum_{\pi,j}\int\frac{d^{2}\xi}{\pi}e^{-|\xi|^{2}}\times\\
    \nonumber &\left\{g^{*}{(J^{+})^{\mu}}_{\nu}\left[\left\{\partial_{\xi^{*}}\psi^{\nu}(t,\xi^{*})\right\}\psi^{*}_{\mu}(t,\xi)-\xi\psi^{\nu}(t,\xi^{*})\psi^{*}_{\mu}(t,\xi)\right]\right.\\
    \label{cond2}&\left.+g{(J^{-})^{\mu}}_{\nu}\left[\xi^{*}\psi^{\nu}(t,\xi^{*})\psi^{*}_{\mu}(t,\xi)-\psi^{\nu}(t,\xi^{*})\left\{\partial_{\xi}\psi^{*}_{\mu}(t,\xi)\right\}\right]\right\}
\end{align}
noting that:
\begin{equation}
    -\partial_{\xi^{*}}e^{-|\xi|^{2}}=\xi e^{-|\xi|^{2}}, \quad -\partial_{\xi}e^{|\xi|^{2}}=\xi^{*}e^{-|\xi|^{2}}
\end{equation}
we can apply integration by parts to terms on the right hand side with derivatives, the boundary terms vanish because of the Gaussian weight $e^{-|\xi|^{2}}$, therefore Eq.~(\ref{cond2}) reduces to:
\begin{align}
    \nonumber &i\frac{d}{dt}\braket{\psi(t)}\\
    \nonumber &=\sum_{\pi,j}\int\frac{d^{2}\xi}{\pi}e^{-|\xi|^{2}}\times \\
    \nonumber &\left\{g^{*}{(J^{+})^{\mu}}_{\nu}(\xi\psi^{\nu}(t,\xi^{*})\psi^{*}_{\mu}(t,\xi)-\xi\psi^{\nu}(t,\xi^{*})\psi^{*}_{\mu}(t,\xi))\right.\\
    &\left.+g{(J^{-})^{\mu}}_{\nu}(\xi^{*}\psi^{\nu}(t,\xi^{*})\psi^{*}_{\mu}(t,\xi)-\psi^{\nu}(t,\xi^{*})\xi^{*}\psi^{*}_{\mu}(t,\xi))\right\}=0\quad .
\end{align}
Thus, for the states evolving according to the equation (\ref{se1}) the probability density remains invariant in time. If we start with a normalized state then following the preceding arguments it should remain normalized throughout its evolution. This establishes our initial claim. 
\end{document}